\documentstyle[12pt,preprint]{aastex}

\shorttitle{Coronagraph Masks}
\shortauthors{Kuchner, Crepp \& Ge}
\begin{document}

\slugcomment{Revised for ApJ 3/7/05.}

\title{Eighth-Order Image Masks For Terrestrial Planet Finding}

\author{Marc J. Kuchner\altaffilmark{1}}
\affil{Princeton University Observatory \\ Peyton Hall, Princeton, NJ 08544}
\altaffiltext{1}{Hubble Fellow}
\email{mkuchner@astro.princeton.edu}

\author{Justin Crepp and Jian Ge}
\affil{Astronomy Department, University of Florida \\
211 Bryant Space Science Center, P.O. Box 112055 \\
Gainesville, FL 32611-2055}
\email{jcrepp@astro.ufl.edu, jge@astro.ufl.edu}

\begin{abstract}

We describe a new series of band-limited image masks for coronagraphy that are insensitive to
pointing errors and other low-spatial-frequency optical aberrations.  For a modest cost in
throughput, these ``eighth-order" masks would allow
the Terrestrial Planet Finder Coronagraph (TPF-C) to operate with a pointing accuracy no better
than that of the Hubble Space Telescope.  We also provide
eighth-order notch filter masks that offer the same robustness to pointing errors combined
with more manageable construction tolerances:  binary masks and sampled graded masks
with moderate optical density requirements.
\end{abstract}


\section{INTRODUCTION}

Coronagraphy holds great promise for imaging extrasolar planetary systems, even
extrasolar terrestrial planets only $\sim 10^{-10}$ times as bright as their host stars \citep[e.g.][]{ks03}.
However, finding extrasolar terrestrial planets at
contrast levels of $\sim 10^{-10}$ using any of the present image mask designs requires either
pointing accuracies at the level of a fraction of a milliarcsecond \citep{kuch02, kuch03}
or apodization in the pupil plane \citep[e.g.][]{aime02,kasd03}, which generally carries a high penalty in throughput
and inner working angle \citep[but see also][]{guyo03, trau03}.
We offer a new series of band-limited image masks that can provide high contrast levels
without pupil apodization because they are intrinsically insensitive to pointing errors and other
low-order aberrations.  We also provide notch filter versions of these masks that
may be easier to build to the necessary tolerances.


\section{EIGHTH-ORDER MASKS}

\subsection{Band-limited Masks and Notch Filter Masks}

Here we summarize the basic definitions of band-limited masks and notch filter masks stated by \citet{kuch02, kuch03}.
We will focus on linear masks, described by functions of a single variable, $x$.
One-dimensional band-limited and notch filter masks can be combined to
create a wide variety of two-dimensional masks.

An ideal linear image mask can be described by a function, $\hat M(x)$, called the mask function.
In our simple model of the interaction between masks and light, the mask function, also called the mask's amplitude transmissivity, multiplies the
electric field phasor of the incoming beam.  The intensity
transmissivity of a mask, $|\hat M({ x})|^2$, multiplies the intensity of the beam.
We will also refer often to the mask function, the mask intensity transmissivity, and to the Fourier
transform of the mask function,
\begin{equation}
M(u)=\int \hat M(x) \, e^{-2 \pi iux} \, dx.
\end{equation}

\citet{kuch02} showed that if $\hat M({x})$ is a notch filter function, i.e.,
\begin{equation}
M(u) = 0 \quad \mbox{for} \,\, \epsilon/2 < |u| < 1 - \epsilon/2,
\label{eq:minusone}
\end{equation}
where $\epsilon$ sets the undersizing of the Lyot stop, and if
\begin{equation}
\int_{-\epsilon/2}^{\epsilon/2} M(u) \, du =0,
\label{eq:zero}
\end{equation}
then the mask defined by $\hat M(x)$ will completely remove all on-axis light in an ideal
coronagraph with a uniform entrance pupil.
\citet{kuch04} showed that notch filter masks are the only trivially achromatic masks
that completely remove on-axis light
in a one-dimensional or separable two-dimensional coronagraph.
Masks we can construct without amplifying the beam or manipulating its phase are necessarily
limited to $0 \le \hat M(x) \le 1$.
A band-limited mask is a notch filter mask with $M(u) = 0$ for $|u| > \epsilon/2$.

We aim to find notch filter mask functions, $\hat M(x)$, that provide deep suppression of
light near the optical axis, not just at the optical axis.  We will
first derive new band-limited masks and then follow the recipes in \citet{kuch03} to
generate useful notch-filter masks based on them.

\subsection{Blocking Slightly-Off-Axis Light}

Understanding the off-axis behavior of an ideal coronagraph with a band-limited mask is easy.
A coronagraph with a band-limited image mask attenuates the intensity of an image of a point source
located at an angle $x$ by a factor of $|\hat M(x)|^2$ compared to the image the source
would have if the image mask were removed while the Lyot stop remained in place (see the Appendix).
In an ideal coronagraph with a band-limited mask, the point spread function (PSF)
is independent of the position of the source with respect to the optical axis;
only the attenuation varies with $x$.  Hence, we can describe the way a band-limited mask
attenuates sources near the optical axis, including the target star, simply by expanding
$\hat M(x)$ about $x=0$.

If the first important term in this expansion is quadratic in $x$, the intensity attenuation will vary as $x^4$.
Borrowing the language of interferometry, we might say such a mask produces a fourth-order
null.  For a demonstration of why this interferometric terminology is
appropriate, consider the nulling coronagraph described by \citet{levi03}, which
monochromatically synthesizes a particular band-limited mask with a fourth-order null
using beam combiners.

All of the band-limited mask designs and notch filter mask designs illustrated in \citet{kuch02},
\citet{kuch03}, and \citet{kuch04} have fourth-order nulls.
For example, all of the popular $1 - \mbox{sinc}^n$ family of masks are fourth order;
$1 - \mbox{sinc}^n k_1 x/n \approx (1/6n) (k_1 x)^2$.  But we can
design band-limited masks and notch filter masks with nulls of any order, $\beta$, by the
methods described below, if $\beta$ is a multiple of 4.

The order of the null dictates the sensitivity of the mask to optical aberrations
that effectively spread the light from a target source around some region of the sky near the optical axis.
Tip-tilt error (caused by pointing error, for example) is the simplest low-order aberration
for us to model and a term that can easily dominate a coronagraph design's error budget.
A pointing error of $\Delta \theta$ will
cause an intensity leak proportional to $(\Delta \theta)^{\beta}$.
A mask that is insensitive
to pointing error will also defeat other low-order aberrations like defocus, coma and
astigmatism to some degree, though some low-order Zernike terms contain mid-spatial-frequency tails
that may leak through.  Mid-spatial-frequency errors are problematic for any
coronagraph design because by definition they coincide with the search area; no mask or
stop can block them without also blocking light from the planet.  \citet{shaklan} discuss
the effects of low-order aberrations in a coronagraph with an eighth-order mask in detail.


The fractional leakage through a mis-pointed coronagraph with a band-limited mask is simply
\begin{equation}
L =  {{\int \int I(x+\Delta \theta,y) \, |\hat M(x,y)|^2 \, dx dy} \over { \int \int I(x,y) \, dx dy}},
\end{equation}
where $I(x,y)$ is
the source intensity, i.e., the stellar disk, and $\Delta \theta$ is the instantaneous pointing error.
For a fourth-order linear mask,
the instantaneous fractional intensity leakage is
\begin{equation}
L = {{\theta_{*}^4 + 48 \, \theta_{*}^2 (\Delta \theta)^2 + 128 (\Delta \theta)^4} \over {256 \, \theta_{IW}^4}},
\label{eq:fourthleak}
\end{equation}
where $\theta_{*}$ is the angular diameter of the star and $\theta_{IW}$ is the inner working angle of the mask,
defined by $|\hat M(\theta_{IW})|^2 =1/2$.  To derive this expression, we made the
approximation that $\hat M(x)=x^4$; we have corrected a numerical error in Equation~17 of \citet{kuch03}.
If we assume $\Delta \theta$ is distributed in a Gaussian with standard deviation
$\sigma_{\Delta \theta}$, and $\sigma_{\Delta \theta} >> \theta_{*}$,
then we find that the mean leakage is
\begin{equation}
\langle L \rangle = 1.5 (\sigma_{\Delta \theta}/\theta_{IW})^4.
\end{equation}
So if we assume that we can tolerate a leakage of $\langle L \rangle < 3 \times 10^{-8}$,
and that $\Delta \theta$ is much larger than the
angular radius of the star, we find that we must center the star on the mask to an accuracy of
\begin{equation}
\sigma_{\Delta \theta} < 0.012 \, \theta_{IW}.
\end{equation}
Though it is easiest to interpret in terms of pointing error, this Gaussian blurring can also serve as
a crude model of the effects of other low-order aberrations.

For an eighth-order mask approximated as $\hat M(x)=x^8$, the instantaneous fractional intensity leakage is
\begin{equation}
L = {{7 \, \theta_{*}^8 + 1120 \, \theta_{*}^6 (\Delta \theta)^2 + 17920 \, \theta_{*}^4 (\Delta \theta)^4 + 57344 \, \theta_{*}^2 (\Delta \theta)^6 + 32768 (\Delta \theta)^8} \over {65536 \, \theta_{1/2}^8}},
\label{eq:eighthleak}
\end{equation}
the corresponding mean fractional leakage is
\begin{equation}
\langle L \rangle = 52.5 (\sigma_{\Delta \theta}/\theta_{1/2})^8,
\end{equation}
and the pointing requirement for leakage $\langle L \rangle < 3 \times 10^{-8}$ is
\begin{equation}
\sigma_{\Delta \theta} < 0.070 \, \theta_{IW},
\end{equation}
a factor of $\sim 6$ improvement over the $\sigma_{\Delta \theta}$ tolerance for fourth-order masks.
A coronagraph designed to find extrasolar terrestrial planets like the
Terrestrial Planet Finder Coronagraph (TPF-C) might need
$\theta_{IW}=60$~milliarcseconds (mas).  This requirement implies a pointing tolerance of
$\sigma_{\Delta \theta} \le 0.72$ mas using a fourth-order mask or $\sigma_{\Delta \theta} \le 4.2$ mas
using an eighth-order mask.  For comparison, the Hubble Space Telescope points to
$\sigma_{\Delta \theta} \approx 3$ mas \citep{burr91}.

Eighth-order masks can also provide high-contrast images of extended sources, though
relaxing the pointing tolerance depletes some of this power.
For a fourth-order mask, Equation~\ref{eq:fourthleak} shows that the extent of a central source
begins to matter when $\theta_{*} > (8/3) (\Delta \theta)$ and the cross term begins to dominate.
For an eighth-order mask,  Equation~\ref{eq:eighthleak} shows that the extent of
a central source begins to be important when $\theta_{*} > (4/7) (\Delta \theta)$.
In the TPF-C example above, these limits correspond to $\theta_{*}=1.9$~mas for a fourth-order mask
and $\theta_{*}=2.4$~mas for an eighth-order mask; a solar-type star at 10 pc is about 1 mas in diameter.
In other words, a TPF-C design with an eighth-order mask may be slightly
better suited for the closest target stars than one using with a fourth-order mask
even with its relaxed pointing tolerance,
depending on the wings of the actual distribution of pointing errors.


\section{CONSTRUCTING THE MASKS}

To design an eighth-order band-limited mask, we can create a linear combination of two fourth-order
band-limited masks weighted so that the term responsible for the quadratic leak cancels; i.e.,
\begin{equation}
\left. {{d^2} \over {dx^2}} \hat M(x) \right|_{x=0} = 0.
\label{eq:cancel}
\end{equation}
For example, we can add a term of the form $C(1 - \cos(k_2 x))$, otherwise known as a $\sin^2$ mask, to
any $1 - \mbox{sinc}^n$ mask to create a new mask with $d^2 \hat M(x)/dx^2 |_{x=0}$, while still
satisfying Equation~\ref{eq:zero}.  If we start with a mask of the form
$1 - \mbox{sinc}^n k_1 x/n \approx (k_1 x)^2/(6n)$
and add $C(1 - \cos(k_2 x)) \approx (C/2) (k_2 x)^2$, we find that to produce an
eighth order mask, we require that $C = -(1/3n) (k_1/k_2)^2$.

However, we do not want to add a $\sin^2$ mask of just any random spatial frequency.  We would prefer a
frequency within the bandwidth of the original mask so that we don't suffer an
undue throughput penalty; i.e., $k_2$ needs to be $\le k_1$.  In order to minimize $|C|$,
we should pick a frequency at exactly the edge of the band; i.e., $k_2 = k_1$.
With this constraint, we find $C = -1/(3 n)$.



Of course, adding mask functions can violate the requirement that $\hat M(x) \le 1$.
To ensure $\hat M(x) \le 1$, we can renormalize the mask by multiplying $\hat M(x)$ by
a constant, $N$, equal to the inverse of its maximum value.

\begin{figure}
\epsscale{0.8}
\plotone{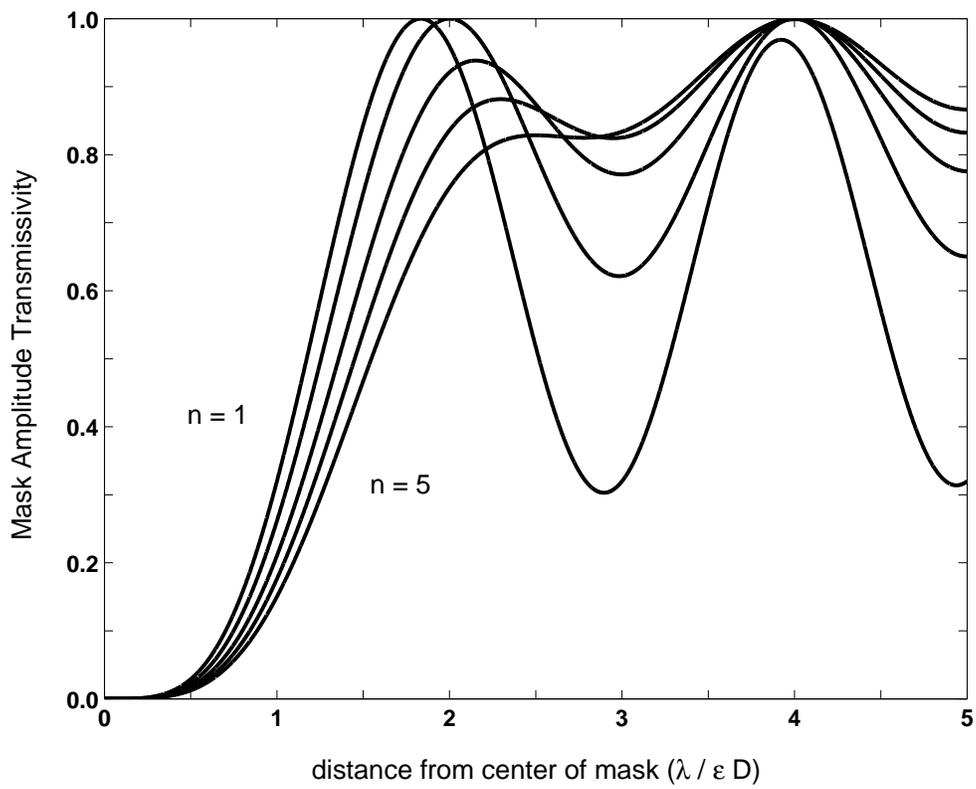}
\caption{Eighth-order band-limited mask functions described by Equation~\ref{eq:eighth} for $n = 1-5$.}
\label{fig:eighthorder}
\end{figure}

Putting everything together and using physical units yields a series of eighth-order band-limited masks,
\begin{equation}
\hat M_{BL}(x) = N \left[\frac{3n-1}{3n} - \mbox{sinc}^n {{\pi x \epsilon} \over {n \lambda_{max} f }}
+ \left({1 \over {3 n}} \right) \cos {{\pi x \epsilon} \over {\lambda_{max} f}} \right],
\label{eq:eighth}
\end{equation}
where $f$ is the focal ratio at the mask and $\lambda_{max}$ is the longest wavelength at which the mask
is to operate. Figure~\ref{fig:eighthorder} shows $\hat M(x)$ for the first few linear masks
in the series. The $n=3$ design offers a good compromise between the large sidelobes of the $n=1$ mask and the
higher inner working angle-bandwidth product of the $n=5$.


The ringing in these image masks reduces their effective throughputs.
The amplitude of the additional ringing introduced by the cosine term in Equation~\ref{eq:eighth}
falls off slowly with $n$, so simply increasing $n$ does not help much.

\begin{figure}
\epsscale{0.8}
\plotone{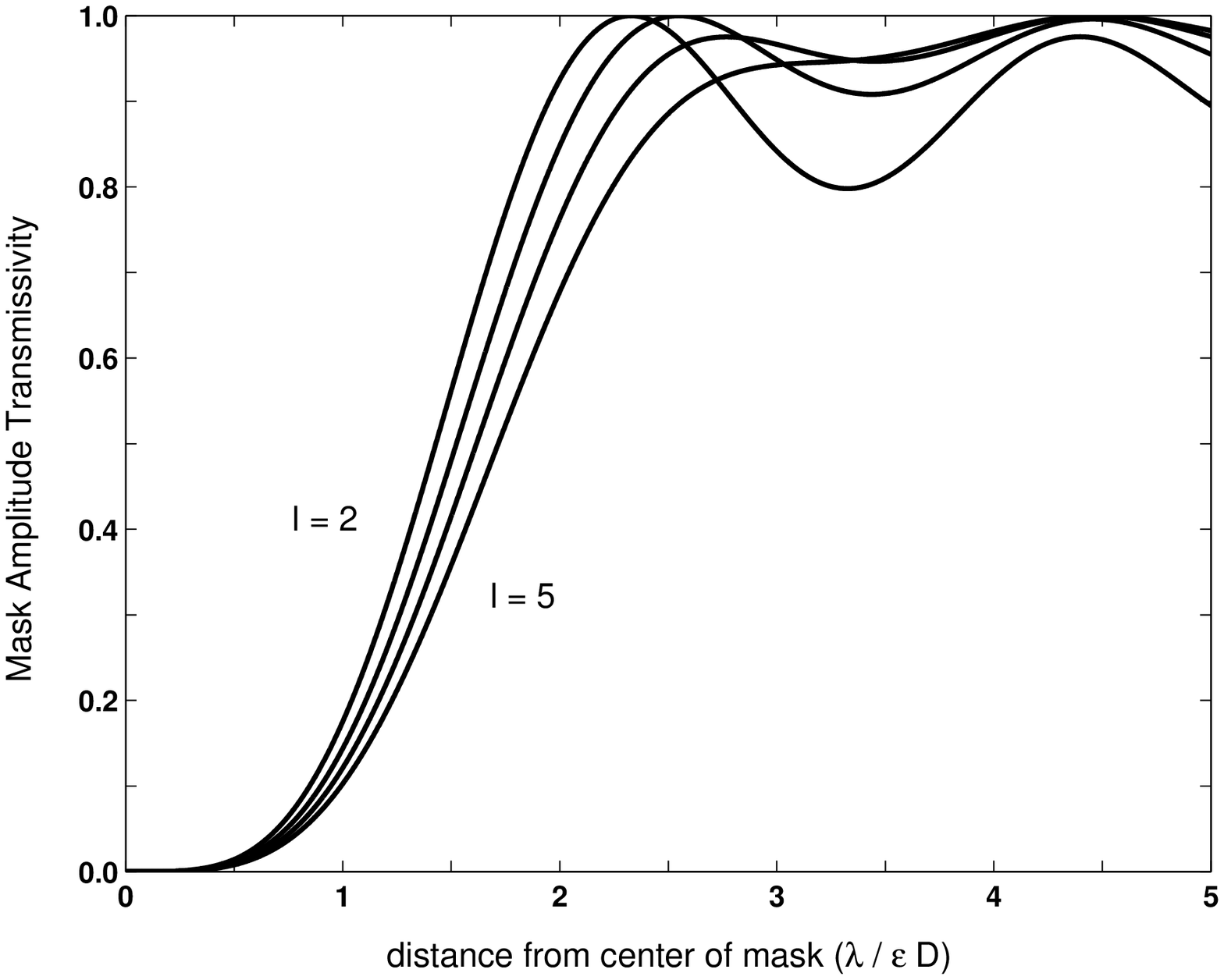}
\caption{Eighth-order band-limited mask functions described by Equation~\ref{eq:eighth2} for $m=1$, $l = 2-5$.}
\label{fig:lowringing}
\end{figure}

Fortunately,
we can create another series of eighth-order masks with less ringing by
combining two $1-\mbox{sinc}^n$ masks instead of a $1 - \mbox{sinc}^n$ mask and $\sin^2$ mask
using the same procedure we used to construct Equation~\ref{eq:eighth}:
\begin{equation}
\hat M_{BL}(x) = N \left[\frac{l - m}{l} - \mbox{sinc}^{l} {{\pi x \epsilon} \over {l \lambda_{max} f}}
+ \frac{m}{l} \, \mbox{sinc}^m {{\pi x \epsilon} \over {m \lambda_{max} f}} \right].
\label{eq:eighth2}
\end{equation}
This series of masks has less ringing than the series described by Equation~\ref{eq:eighth}.
It is parametrized by two integer exponents, $l$ and $m$; we assume $l > m$.
Figure~\ref{fig:lowringing} shows $\hat M(x)$ for $m = 1$ and
$l =$2--5. The $m=1$ and $l=$2--3 masks have throughput similar to the $n = 3$ cosine mask. Using
large values of $m$ and $l$ reduces the ringing further, but it also reduces the Lyot stop throughput.

\begin{figure}
\epsscale{0.8}
\plotone{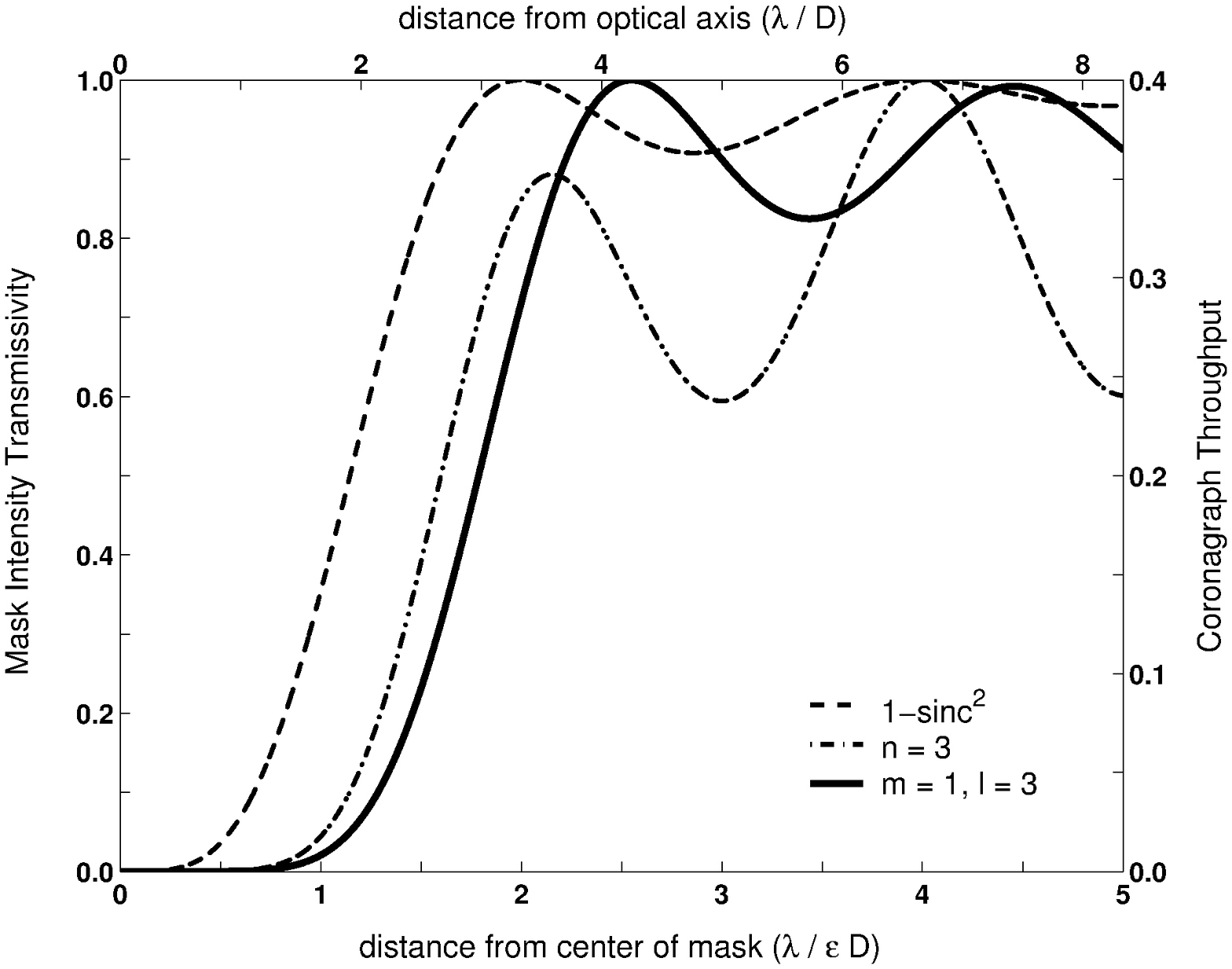}
\caption{Intensity transmissivities for the $1- \mbox{sinc}^2$ fourth-order mask, the $n=3$ eighth-order mask, and the $m=1$,
$l=3$ eighth-order mask. Coronagraph throughput and distance from optical axis were calculated with $\epsilon=0.6$.
The $m=1$, $l=3$ eighth-order mask, recommended for TPF-C, has 100\% transmissivity at its first maximum.}
\label{fig:itf}
\end{figure}

Figure~\ref{fig:itf} compares the intensity transmissivity, $|\hat M(x)|^2$, for the $1 - \mbox{sinc}^2$ fourth-order mask
and the $m=1$, $l=3$ eighth-order mask.  While the $1-\mbox{sinc}^2$ mask has an inner working angle of
$\theta_{IW} = (1.448/\epsilon)(\lambda/D)$, the $m=1$, $l=3$ eighth-order mask has an inner working angle of $\theta_{IW} = (1.788/\epsilon)
(\lambda/D)$.   The $m=1$, $l = 3$ mask offers a good compromise between ringing and throughput, and also reaches 100\% transmissivity at
its first maximum, a critical region for planet searching; we recommend this mask for TPF-C.


Consider a TPF-C design with $\theta_{IW} = 3 \, \lambda/D$ using a linear fourth-order mask.  This
coronagraph has a bandwidth of $\epsilon =0.4$ and a nominal Lyot stop throughput of $1-\epsilon = 0.6$ \citep[see][]{kuch03}.
This fourth-order design probably requires some mild apodization of the Lyot stop
to ameliorate leakage due to low-order optical aberrations, reducing the
throughput to $0.35$. Keeping $\theta_{IW} = 3 \, \lambda/D$ but switching to a linear
$m=1$, $l=3$ eighth-order mask would mean working at a bandwidth of $\epsilon=0.596$
and a Lyot stop throughput of $1-\epsilon=0.404$.  Coronagraphs with eighth-order masks should
not require any Lyot stop apodization.


In other words, our analysis suggests that eighth-order masks combined with un-apodized Lyot stops
perform about as well as fourth-order masks combined with apodized Lyot stops in terms of
throughput and robustness to pointing errors.  An alternative way to provide robustness
to pointing errors is to use a shaped-pupil coronagraph \citep{kasd03, vand03a, vand03b}.
But the throughput offered by an eighth-order linear mask is still better than the typical throughput
of a shaped pupil coronagraph at a given inner working angle, particularly when a shaped pupil coronagraph
is used with a hard-edged image mask, which increases its effective inner working angle \citep{kuch04}.

\section{EIGHTH-ORDER NOTCH FILTER MASKS}
\label{sec:binarymasks}

The functions described by Equations~\ref{eq:eighth} and \ref{eq:eighth2} can be used in a variety of ways, e.g., to make linear masks
($\hat M(x,y) = \hat M_{BL}(x)$), radial masks ($\hat M(r) = \hat M_{BL}(r)$), or separable masks
($\hat M(x,y) = \hat M_{BL}(x) \hat M_{BL}(y)$).  However, all band-limited masks are necessarily smooth graded masks.
Notch-filter masks offer even more design freedom and need not necessarily be smooth, making them potentially easier to manufacture than
band-limited masks \citep{kuch03}.

Notch filter masks affect starlight and planet-light the same way as band-limited masks;
only their low-spatial frequency parts contribute to starlight suppression.
Consequently, in an eighth-order notch filter mask, only the low-frequency part
needs to satisfy Equation~\ref{eq:cancel}.  In other words,
\begin{equation}
\left. {{d^2} \over {dx^2}} \left( \mbox{the low-frequency part of} \ \hat M(x) \right) \right|_{x=0}
= \int_{-\epsilon/2}^{\epsilon/2} (-2 \pi i u)^2 M(u) \, du = 0.
\end{equation}
Equivalently, we can say that an eighth-order notch filter mask satisfies Equations~\ref{eq:minusone}, \ref{eq:zero}, and also
\begin{equation}
\int_{-\epsilon/2}^{\epsilon/2} u^2 M(u) \, du =0.
\label{eq:zero2}
\end{equation}

To find masks that meet these criteria, we can use a technique similar to the one employed in $\S$3.
Since any linear combination of fourth-order notch filter mask functions will automatically satisfy
Equations~\ref{eq:minusone} and \ref{eq:zero}, we will start by writing
\begin{equation}
\hat M_{notch_8}(x)=N\left[ \hat M_{notch_{4A}}(x)+ C \hat M_{notch_{4B}}(x) \right],
\label{eq:8notch}
\end{equation}
where $\hat M_{notch_{4A}}(x)$ and $\hat M_{notch_{4B}}(x)$ represent different fourth-order
notch filter mask functions and $N$ ensures that $\hat M_{notch_8}(x) \leq 1$.  To construct
a notch filter mask that exhibits eighth-order behavior, we need to weight the linear combination so that
the new notch filter function also satisfies Equation~\ref{eq:zero2}.  In other words we will find
the constant $C$ by substituting Equation~\ref{eq:8notch} into~\ref{eq:zero2}.
\begin{equation}
C=-\frac{\int_{-\epsilon/2}^{\epsilon/2} u^2 M_{notch_{4A}}(u) \, du}{\int_{-\epsilon/2}^{\epsilon/2} u^2 M_{notch_{4B}}(u) \, du}.
\label{eq:C1}
\end{equation}
This constant should be negative; it should also satisfy $|C|<1$


By combining fourth-order notch filter functions and using the solutions to Equation~\ref{eq:C1},
we can construct a variety of eighth-order notch filter masks analogous to the variety of eighth-order
band-limited masks.  For example, we can make a family of eighth-order notch filter masks using the
$1-\mbox{cos}$ and $1-\mbox{sinc}^n$ fourth-order notch filter functions. We can also design
low-ringing eighth-order notch filter masks using the $1-\mbox{sinc}^m$ and $1-\mbox{sinc}^l$ notch
filter functions.  To be consistent with $\S$3, we will refer to the various eighth-order notch filter
masks by the exponents of their constituent functions ($n$, $m$, $l$, ... etc.). In the following,
we provide example calculations for making eighth-order binary and graded notch filter masks
using the $m=1$, $l=3$ design.

\subsection{Eighth-order Binary Masks}

Notch filter masks can be designed to be binary: everywhere either completely opaque or completely transparent. A simple way to make such a binary
mask is to assemble a mask from a collection of identical parallel stripes, where any arbitrary notch filter mask function provides the width of each
stripe. In other words, each stripe is defined by
\begin{equation}
\hat M_{stripe}(x,y)=
\left\{
\begin{array}{ll}
1  & \mbox{where} \,\, y < \hat M_{notch}(x)\, \lambda_{min} f \\
0  & \mbox{elsewhere}.
\end{array}
\right.
\label{eq:stripe}
\end{equation}
and the mask function is
\begin{equation}
\hat M_{binary}(x,y) = \sum_{j=-\infty}^{\infty} \hat M_{stripe}(x,\;y-j \lambda_{min} f),
\label{eq:stripedmask}
\end{equation}
where $\lambda_{min}$ is the shortest wavelength in the band of interest.

If we like, we can use the band-limited mask functions described by Equations~\ref{eq:eighth} or~\ref{eq:eighth2}
in place of $\hat M_{notch}(x)$, resulting in a mask formed of continuous curves.  However, sampled binary masks may prove to
be easier to manufacture since their features are not as small near the optical axis. We will construct
here an eighth-order sampled binary mask. Such a mask can be made entirely from rectangles of opaque material.
\citet{debe04} have demonstrated the construction of sampled fourth-order masks using e-beam lithography.

Fourth-order sampled masks are defined by the following prescription \citep{kuch03}:
\begin{equation}
\hat M_{notch_4}(x)=\hat M_{samp_4}(x) - \hat M_0,
\label{eq:sampminusm0}
\end{equation}
where
\begin{mathletters}
\begin{eqnarray}
\hat M_{samp_4}(x)
&=&\hat P(x) * \left(\hat M_{BL_4}(x) \,{\Delta x} \sum_k \delta(x-(k+\zeta)\Delta x)\right),  \\
M_{samp_4}(u)
&=&P(u) \left(M_{BL_4}(u) * \sum_k \delta(u-k/\Delta x) e^{-2 \pi i u \zeta \Delta x} \right),
\label{eq:tsampled}
\end{eqnarray}
\end{mathletters}
and
\begin{equation}
\hat M_0=\int_{-\epsilon D/(2 \lambda)}^{\epsilon D/(2 \lambda)} M_{samp_4}(u) \, du
=\int_{-\infty}^{\infty} M_{BL_4}(u) P(u) \, du
=\left. \hat M_{BL_4}(x) * \hat P(x) \right|_{x=0}.
\label{eq:m0}
\end{equation}
Here, $M_{BL_4}$ represents any fouth-order band-limited mask function, $k$ ranges over all integers, and
$*$ indicates convolution.  The sampling points are offset from the mask center by a fraction $\zeta$ of $\Delta x$.
The kernel, $\hat P(x)$, can represent the ``beam'' of a nanofabrication tool.  It
should be normalized so that $\int_{-\infty}^{\infty} \hat P(x) \, dx = 1$,
and $\hat P(x)$ must be everywhere $\le 1/(\Delta x)$, so
$\hat M_{samp_4}(x)$ remains $\le 1$. The constant $\hat M_0$ ensures that the mask
satisfies Equation~\ref{eq:zero}.  Though the sampled mask is derived from $\hat M_{BL_4}(x)$,
the function being sampled is $\hat M_{BL_4}(x)  - \hat M_0$.

Combining Equation~\ref{eq:8notch} and Equation~\ref{eq:sampminusm0}, we have
\begin{eqnarray}
\hat M_{notch_8}(x) &=& N\left[ \hat M_{notch_{4A}}(x) + C \hat M_{notch_{4B}}(x) \right] \\
                  &=& N\left[ ( \hat M_{samp_{4A}}(x) - \hat M_{0_A} ) + C ( \hat M_{samp_{4B}}(x) - \hat M_{0_B} ) \right],
\label{eq:bump}
\end{eqnarray}
where $\hat M_{samp_{4A,B}}(x)$
are sampled versions of the fourth-order band-limited functions $\hat M_{BL_{4A,B}}(x)$
described by Equation~\ref{eq:tsampled}.   The constants  $\hat M_{0_{A,B}}$ and $C$
ensure that $\hat M_{notch_{4A,B}}(x)$ satisfy both Equation~\ref{eq:zero} and Equation~\ref{eq:zero2},
The constant $C$ is given by
\begin{equation}
C=-\frac{\int_{-\epsilon/2}^{\epsilon/2} u^2 P(u) M_{BL_{4A}}(u) \, du}{\int_{-\epsilon/2}^{\epsilon/2} u^2 P(u) M_{BL_{4B}}(u) \, du}.
\label{eq:C}
\end{equation}
To make an eighth-order sampled notch filter mask, the function that we sample is
$N[(\hat M_{BL_{4A}}(x)-\hat M_{0_A})+ C(\hat M_{BL_{4B}}(x)-\hat M_{0_B})]$.



\begin{figure}
\epsscale{0.8}
\plotone{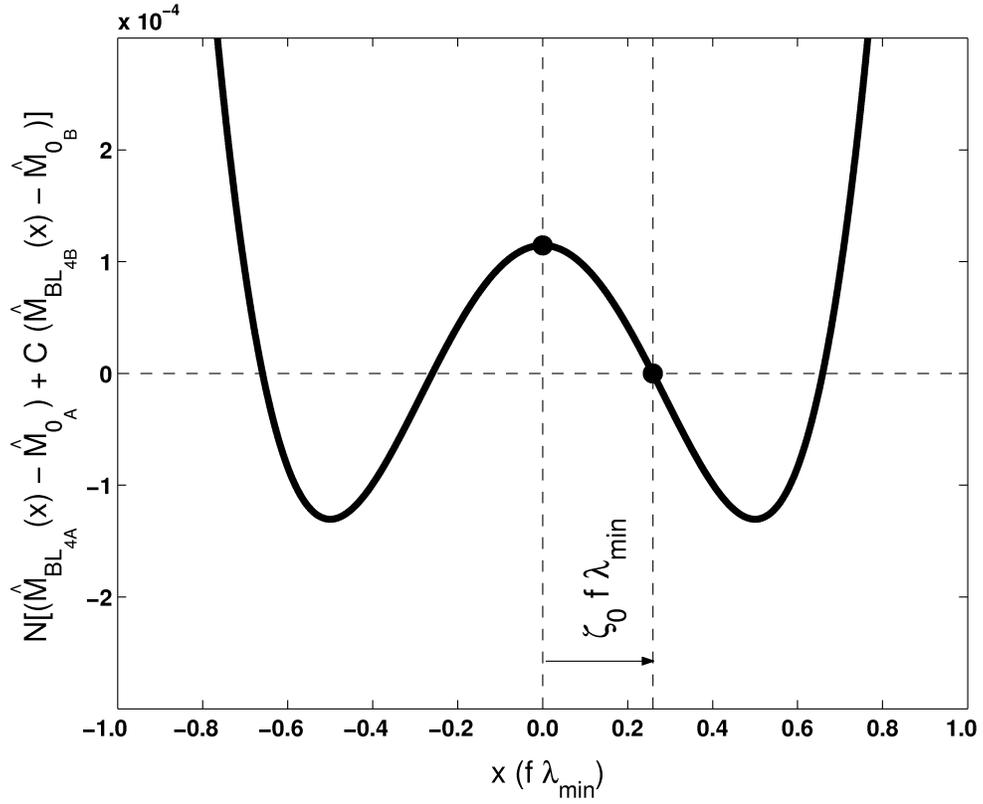}
\caption{An example of the function $N[(\hat M_{BL_{4A}}(x)-\hat M_{0_A})+C(\hat M_{BL_{4B}}(x)-\hat M_{0_B})]$ for an
$m=1$, $l=3$ sampled mask.  Choosing $\zeta=\zeta_0$ allows us to create a binary mask of contiguous stripes.  Choosing $\zeta=0$ allows us to
create a graded mask with the most favorable optical density requirement.}
\label{fig:samp}
\end{figure}

Figure~\ref{fig:samp} shows a plot of the function
$N[(\hat M_{BL_{4A}}(x)-\hat M_{0_A})+ C(\hat M_{BL_{4B}}(x)-\hat M_{0_B})]$
to illustrate how $\zeta$ may be chosen.
This example uses the $m=1$, $l=3$ sampled eighth-order mask, meaning
$\hat M_{BL_{4A}}(x)= 1- \mbox{sinc}(x)$ and $\hat M_{BL_{4B}}(x)=1 - \mbox{sinc}^3(x)$.
To guarantee that $\hat M_{notch}(x) \ge 0$, the parameter $\zeta$ must be in the
range $|\zeta| \leq \zeta_0$, where $\zeta_0$ is defined by the condition
$\hat M_{BL_{4A}}(\zeta_0 \lambda_{min} f) + C \hat M_{BL_{4B}}(\zeta_0 \lambda_{min} f) =  \hat M_{0_A} + C \hat M_{0_B}$.
For our binary mask, we will choose $\zeta=\zeta_0$, to make the central
rectangles contiguous.

\begin{figure}[!ht]
\epsscale{0.8}
\centerline{
\includegraphics[height=2.2in]{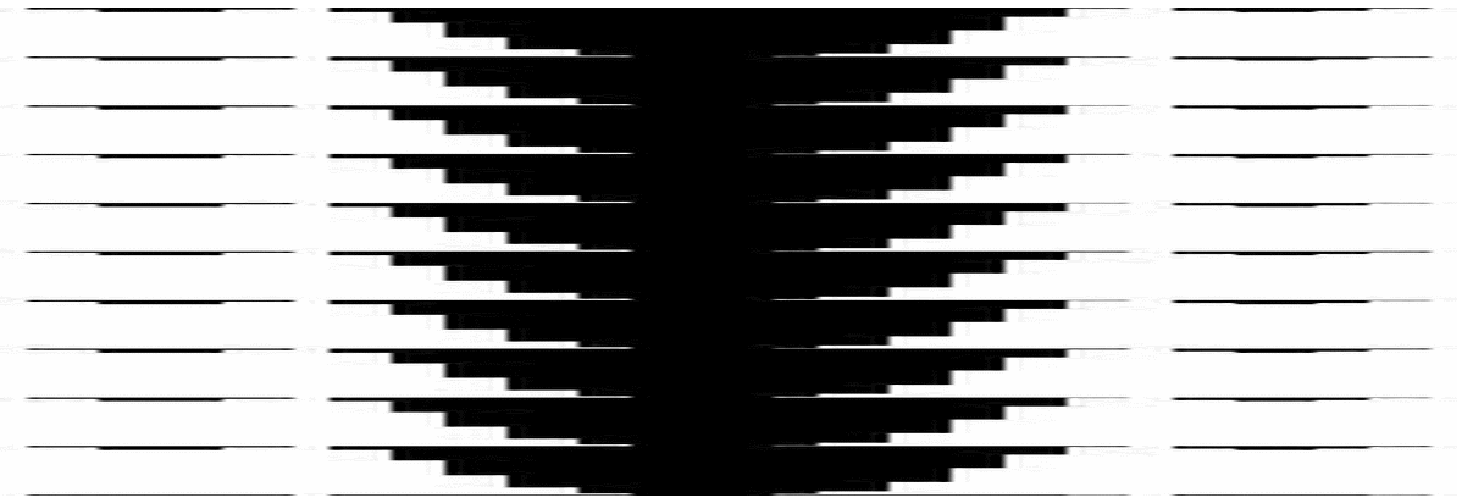}
}
\caption{Simulated high magnification picture of an $m=1$, $l=3$ linear
eighth-order sampled binary mask. Dark areas are completely opaque and white areas perfectly
transmissive.  See \citet{debe04}
for optical microscope photographs of an actual linear fourth-order 1-sinc$^2$ binary mask.}
\label{fig:binary}
\end{figure}

The bandwidth of a mask should be chosen conservatively; e.g., $\lambda_{max}$ should be somewhat larger then the longest wavelength
where the detector is sensitive so a filter with a finite slope can remove all the extraneous light. Band-limited masks and notch filter masks
leak light at wavelengths longer than $\lambda_{max}$; notch filter masks also leak light at wavelengths shorter than $\lambda_{min}$. At a fixed inner working angle,
increasing $\lambda_{max}$ necessitates increasing $\epsilon$, and thereby decreasing the throughput. Decreasing $\lambda_{min}$ means spacing
the stripes and samples in a notch filter mask closer together.

For the $m=1$, $l=3$ mask with $\theta_{IW} = 3 \, \lambda_{max}/D$, spacing $\Delta x = \lambda_{min} f$, and bandpass 0.5--0.8~$\mu$m, we find
that $\hat M_{0_A}=0.00630889$, $\hat M_{0_B}=0.01882618$, $C=-0.33935486$, and $\zeta_0=0.25941279$. Table~\ref{tab:maskparams} lists
normalization constants and sampled mask parameters for eighth-order notch filter masks of various inner-working-angles using a top-hat kernel,
$\hat P(x)=(D/ \lambda_{min}) \Pi (x D/ \lambda_{min})$, and 0.5--0.8~$\mu$m bandpass.

If the resolution of our nanotool is $\sim\;$20 nm, we require a telescope with an f/115 or slower beam \citep[see][]{kuch03}.
The physical size of an entire mask is generally a few hundred diffraction widths. A 1'' $\times$ 1'' mask would consist
of $\gtrsim 440$ vertically repeating segments, where each segment is $\leq \lambda_{min} f = 57.5\;\mu m$ wide. This coronagraph
design would have a Lyot stop throughput of 40\%. Figure~\ref{fig:binary} shows an example of what an $m=1$, $l=3$ linear
eighth-order binary mask would look like.

A similar mask can also be made by replacing Equation~\ref{eq:stripe} with
\begin{equation}
\hat M_{stripe}(x,y)=
\left\{
\begin{array}{ll}
1  & \mbox{where} \,\, |y| < \hat M_{notch}(x)\, \lambda_{min} f/2 \\
0  & \mbox{elsewhere}.
\end{array}
\right.
\label{eq:stripe2}
\end{equation}
as shown in \cite{kuch03}. In this design, the notch-filter function is reflected vertically for each segment.
Manufacturing a mask like the one shown in Figure~\ref{fig:binary} could substantially reduce the writing time
for e-beam lithography and opportunities for write errors.

\begin{deluxetable}{lcccccccc}
\rotate
\tablewidth{8in}
\tablecaption{SAMPLED EIGHTH-ORDER MASK PARAMETERS}
\startdata
\tableline
\tableline
n        &   N$\tablenotemark{a}$  &   $\theta_{IW}$ ($\lambda_{max}/D$) &  $\epsilon$ &  $\hat M_{0_A}$  &  $\hat M_{0_B}$ & $C$ & $\zeta_0$ & O.D.$_{\rm{max}}$$\tablenotemark{b}$ \\
\tableline
1        &  0.966115405054    &    3 &     0.453 &   0.01092315   & 0.01631996  & -0.67562610 & 0.25945500 & 8.004 \\
         &  0.960497496515    &    4 &     0.340 &   0.00616927   & 0.00923329  & -0.67167174 & 0.25954657 & 9.012 \\
         &  0.959860814806    &    5 &     0.272 &   0.00395309   & 0.00592117  & -0.66985758 & 0.25958912 & 9.791 \\
\tableline
2        &  0.999927046667    &    3 &     0.487 &   0.00632078   & 0.01883265  & -0.34113344 & 0.25944228 & 7.969 \\
         &  0.999967637078    &    4 &     0.366 &   0.00357716   & 0.01068997  & -0.33769399 & 0.25953913 & 8.969 \\
         &  0.999991487843    &    5 &     0.292 &   0.00227903   & 0.00682024  & -0.33609563 & 0.25958482 & 9.757 \\
\tableline
3        &  0.994355716928    &    3 &     0.533 &   0.00505061   & 0.02250740  & -0.22964671 & 0.25941276 & 7.860 \\
         &  0.992967898001    &    4 &     0.400 &   0.00284944   & 0.01275232  & -0.22635050 & 0.25952294 & 8.868 \\
         &  0.992249764357    &    5 &     0.320 &   0.00182510   & 0.00818418  & -0.22484879 & 0.25957404 & 9.649 \\
\tableline
4        &  0.999920502046    &    3 &     0.578 &   0.00445606   & 0.02640459  & -0.17384571 & 0.25937869 & 7.744 \\
         &  0.999959029620    &    4 &     0.434 &   0.00251629   & 0.01499184  & -0.17065228 & 0.25950368 & 8.751 \\
         &  1.000006649135    &    5 &     0.347 &   0.00160975   & 0.00961517  & -0.16919646 & 0.25956186 & 9.534 \\
\tableline
\tableline
\multicolumn{2}{l}{$l$ (for $m=1$)}                 &      &           &      &  & & &\\
\tableline
2        &  1.865785172445    &    3   &   0.557  &  0.00825681   & 0.01646433  & -0.50505400 & 0.25942680 & 7.923 \\
         &  1.862096989484    &    4   &   0.412  &  0.00452972   & 0.00904463  & -0.50274761 & 0.25953456 & 8.977 \\
         &  1.856230853161    &    5   &   0.334  &  0.00298028   & 0.00595415  & -0.50180101 & 0.25957920 & 9.710 \\
\tableline
3        &  1.434216871605 & 3$\tablenotemark{c}$ & 0.596 & 0.00630889 & 0.01882618 & -0.33935486 & 0.25941279 & 7.882 \\
         &  1.429552473250    &    4   &   0.447  &  0.00355642   & 0.01063737  & -0.33669458 & 0.25952307 & 8.890 \\
         &  1.427349701514    &    5   &   0.357  &  0.00227076   & 0.00679929  & -0.33546973 & 0.25957440 & 9.674 \\
\tableline
4        &  1.312506672966    &    3   &   0.637  &  0.00540801   & 0.02147409  & -0.25623997 & 0.25623997 & 7.813 \\
         &  1.308947497039    &    4   &   0.478  &  0.00305104   & 0.01215389  & -0.25348152 & 0.25951079 & 8.819 \\
         &  1.306220598720    &    5   &   0.382  &  0.00195033   & 0.00778076  & -0.25221408 & 0.25956647 & 9.603 \\
\tableline
\enddata
\tablenotetext{a}{Normalization constant for $\zeta=\zeta_0$ and $f \lambda_{min}$ sampling.}
\tablenotetext{b}{For a graded mask with $\zeta=0$.}
\tablenotetext{c}{Suggested for TPF-C.}
\label{tab:maskparams}
\end{deluxetable}




\subsection{Eighth-order Graded Masks}
\label{sec:gradedmasks}

Smooth graded band-limited image masks have produced suppression of on-axis monochromatic light at 
the level of a few times $10^{-9}$ in the laboratory \citep{trau04}. However, construction errors 
probably still limit the broad-band performance of these masks. We suggest that sampled graded masks 
may be easier to construct than smooth graded masks. Graded masks produce large phase errors, but it 
may be possible to correct the phase of these sampled masks using transparent strips of varying 
thickness. Also, as \citet{kuch03} pointed out, sampled masks can be designed so that, unlike smooth 
masks, they do not require their darkest regions to be perfectly opaque. This flexibility limits the
demands on the lithography tool used to make the masks. The $1-\mbox{sinc}^2$ mask with
$\theta_{IW} = 2.9 \, \lambda_{max}/D$, $\epsilon=0.4$, can be built with a maximum optical
density of 4. The $\sin^2$ mask with $\epsilon=0.4$ can be built with a maximum optical density of 3.

\begin{figure}[!ht]
\centerline{
\includegraphics[height=2.2in]{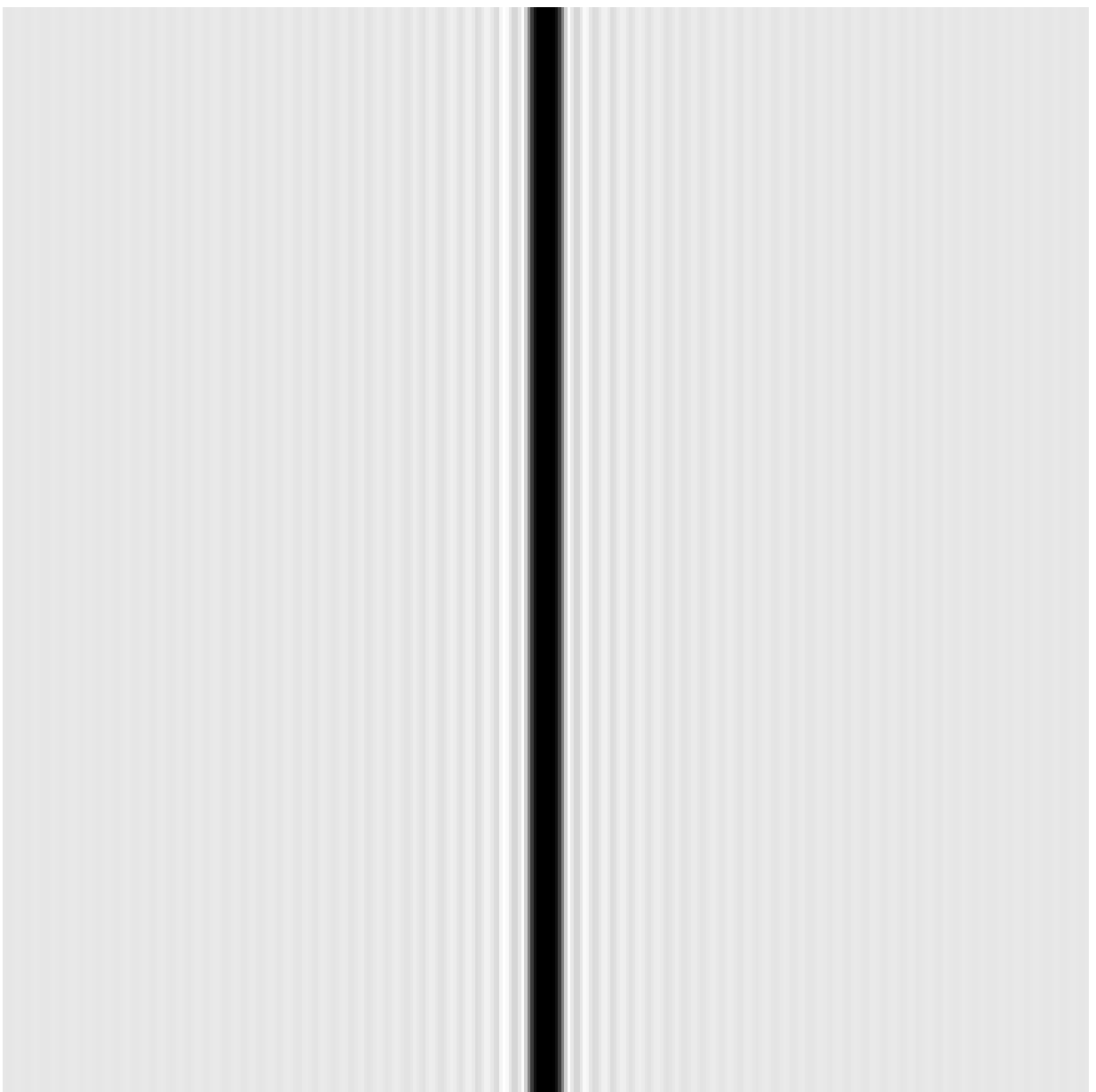}
\hspace{0.8in}
\includegraphics[height=2.09in]{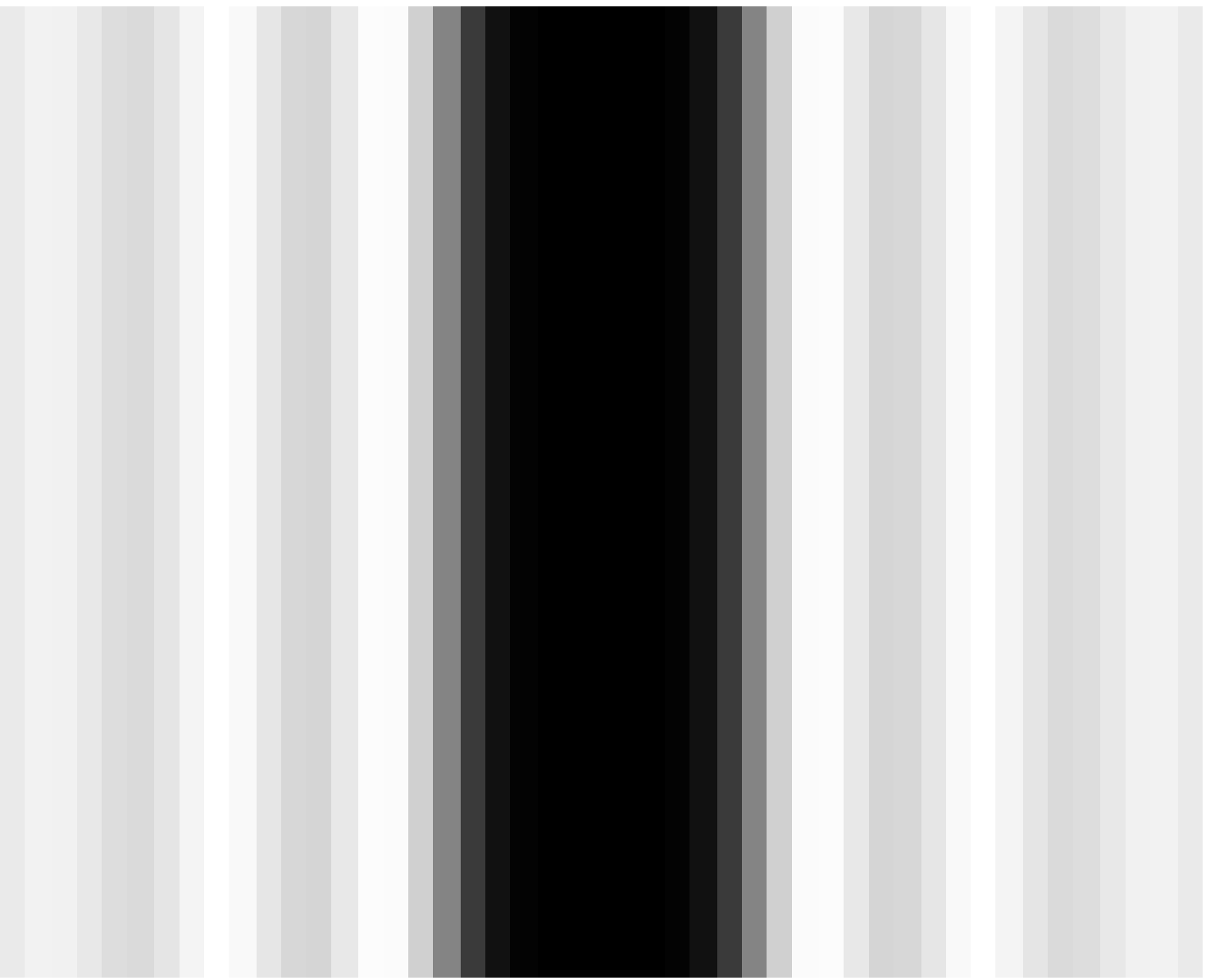}
}
\caption{Simulated low and high magnification pictures of an $m=1$, $l=3$ eighth-order sampled graded mask with $\zeta=0$. The
low magnification picture (left) shows $\sim 400$ diffraction widths; each stripe has uniform shading and is $f \lambda_{min}$ wide.}
\label{fig:graded}
\end{figure}

When we design eighth-order graded notch filter masks, we can reduce the required maximum optical density by
beginning the sampling at $\zeta=0$, so long as the spacing between the samples is large enough to straddle the valleys shown in Figure~\ref{fig:samp}. Choosing
$\Delta x = \lambda_{min} f$ satisfies this condition for all of the masks listed in Table~\ref{tab:maskparams}.
Figure~\ref{fig:graded} shows a graded version of the $m=1$, $l=3$ eighth-order mask described in $\S$4.1. The mask is defined
by $\hat M(x,y) = \hat M_{notch}(x)$; its optical density is $-\log_{10} |\hat M_{notch}(x)|^2$. To make the darkest stripe of
the mask as transparent as possible, we chose $\zeta=0$. With this choice, the darkest stripe of the mask has optical density
$- 2 \log_{10} |\hat M_{notch}(0)| \approx 7.882$. Table~\ref{tab:maskparams} lists the maximum optical densities
(O.D.$_{\rm{max}}$) of sampled graded masks with $\zeta=0$.

\section{SUMMARY}
\label{sec:conclusion}

We offered a series of eighth-order masks that are relatively insensitive
to tip-tilt errors and other low-spatial-frequency aberrations; in a coronagraph using one of these masks,
the r.m.s.~pointing error only needs to be managed to a few milliarcseconds,
no better than the pointing accuracy of the Hubble Space Telescope.
Eighth-order notch filter masks retain most benefits of using
fourth-order masks---broadband capabilities, reasonably high throughput, and small inner
working angle---permitting extremely-high dynamic range coronagraphy suitable for
terrestrial planet finding using a popular optical layout.

In particular, we suggested a binary mask designed for TPF-C at 0.5--0.8~$\mu$m
composed of opaque strips whose shapes are described by
Equation~\ref{eq:bump} with $m=1$, $l=3$, $\epsilon=0.596$,
$N = 1.434216871605$, $\hat M_{0_A}=0.00630889$, $\hat M_{0_B}=0.01882618$,
$C=-0.33935486$, and $\zeta_0=0.25941279$. This mask provides 40\% Lyot
stop throughput and requires an f/115 or slower beam,
assuming the mask can be manufactured with an r.m.s.~accuracy of 20~nm.
The r.m.s.~pointing accuracy required for achieving starlight
suppression of $10^{-10}$ with this mask in the search area is
$\sigma_{\Delta \theta} \approx 4.2$ milliarcseconds
for stars of diameter up to $\sim 2.4$ mas.  If the mask is used on a
telescope with better pointing accuracy, it can
achieve contrast levels of $10^{-10}$ on targets with even larger diameters.

We also provided a graded version of this design, whose amplitude transmissivity is
described by Equation~\ref{eq:bump} using the above parameters but with $\zeta=0$.
This mask offers the same performance as the above binary version, but it allows
easier e-beam fabrication because it only requires optical densities $\leq 7.882$.  Other
eighth-order masks can provide less ringing at the cost of inner working angle
or Lyot stop throughput. 

%

\acknowledgments

We thank Stuart Shaklan and Joseph Green for helpful conversations and for delaying
the publication of their paper on low-order aberrations in coronagraphs with
eighth-order masks until this paper was ready.  M.J.K. acknowledges the support of the
Hubble Fellowship Program of the Space Telescope Science Institute.
J.C. and J.G. acknowledge support by NASA with grants
NAG5-12115, NAG5-11427, NSF with grants AST-0138235 and AST-0243090, the
UCF-UF Space Research Initiative program, and the JPL TPF program.

\appendix

\section{APPENDIX}

We will prove for a monochromatic coronagraph with a notch filter mask, a
binary entrance aperture of finite size, and a Lyot stop that is perfectly opaque
everywhere the entrance aperture is opaque, that
1) the PSF shape is the absolute square of the Fourier transform of the Lyot stop amplitude transmissivity
independent of the position of the source on the sky, and 2) the PSF is attenuated by
the intensity transmissivity of the band-limited part of the
mask evaluated at the source position.  \citet{kuch02} demonstrated this
principle for a $\sin^2$ mask; this more general proof applies to any two-dimensional
notch filter mask.

As usual, we will examine a coronagraph comprising
an entrance aperture, $A$, an image mask, $\hat M$, and a Lyot stop, $L$,
each of which is represented by a complex-valued function.
We will use the notational conventions of \citet{kuch02} and \citet{kuch03}:
letters with hats represent image plane quantities.
The image-plane coordinates are ${\bf x}=(x,y)$
and the pupil-plane coordinates are ${\bf u}=(u,v)$.

Monochromatic light propagates through the coronagraph as follows.
\begin{trivlist}
\item{1) An incoming wave incident on the entrance aperture creates a field with amplitude $E({\bf u})$.
When an incoming wave interacts with a stop or mask, the function
representing the mask multiplies the wave's complex amplitude.  So after the wave interacts with
the entrance aperture, the amplitude becomes $A({\bf u})\cdot E({\bf u})$.}

\item{2) After the entrance aperture, the beam propagates to an image plane, where the new field amplitude is the Fourier transform
of the pupil plane field amplitude, $\hat A({\bf x}) * \hat E({\bf x})$; $*$ denotes convolution. In this plane, the beam interacts with the image mask, and the field amplitude becomes $\hat M({\bf x}) \cdot (\hat A({\bf x}) * \hat E({\bf x}))$.}

\item{3) Next, the beam propagates to a second pupil plane, where the field amplitude is $M({\bf u}) * (A({\bf u}) \cdot  E({\bf u}))$.
In this second pupil plane, the wave interacts with a Lyot stop, changing
the field amplitude to $F({\bf u}) = L({\bf u}) \cdot [M({\bf u}) * (A({\bf u}) \cdot E({\bf u}))]$.}

\item{4) At last, the beam propagates to the final image plane, where the final image field is $\hat F({\bf x})$,
the Fourier transform of $F({\bf u})$.  For a point source, the intensity of the final image is proportional to the
absolute value of this quantity squared.}
\end{trivlist}
The final image field, $\hat F(\bf x)$, and its Fourier transform are linear functions
of $A(\bf u)$, $L(\bf u)$, and also $M(\bf u)$.  This last property allows us to study masks
by decomposing them into Fourier components, computing $F({\bf u})$ or $\hat F(\bf x)$ for each one,
and then summing the final field amplitudes back together.

Consider a point source providing a field $\hat E({\bf x}) = \delta({\bf x}-{\bf x_1})$ in the plane
of the sky and a harmonic mask function $M({\bf u})= \delta({\bf u}-{\bf u_1})$.
The field after the entrance pupil is
$A({\bf u}) \exp(-2 \pi i \, {\bf u} \cdot {\bf x_1})$, and the field in the first image plane is
$\hat A({\bf x}-{\bf x_1})$.  The field after the image mask is $ \exp(2 \pi i  \,{\bf u_1} \cdot {\bf x}) \hat A({\bf x}-{\bf x_1})$.
The field in the second pupil plane is $ A({\bf u}-{\bf u_1}) \exp(-2 \pi i ({\bf u}-{\bf u_1}) \cdot {\bf x_1})$.
The field after the Lyot stop is
\begin{equation}
F({\bf u})=L({\bf u}) A({\bf u}-{\bf u_1}) e^{-2 \pi i ({\bf u}-{\bf u_1}) \cdot {\bf x_1}}
\qquad \mbox{for a harmonic mask.}
\label{eq:fharmonic}
\end{equation}

Let $A$ be binary (everywhere equal to 1 or 0) and let
$\cal A$ represent the support of $A$ and $\cal L$ represent the support of $L$.
If $\cal L \in \cal A$, then there is some set ${\cal P} \in {\cal R}^2$ for which
$L({\bf u}) A({\bf u}-{\bf u_1}) = L({\bf u})$ for $\bf u_1 \in {\cal P}$.
If $\cal A$ is finite in extent, then there is also some set ${\cal Q} \in {\cal R}^2$ for which
$L({\bf u}) A({\bf u}-{\bf u_1}) = 0$ for $\bf u_1 \in {\cal Q}$.

Under these circumstances, there are three kinds of harmonic image masks:
\begin{trivlist}
\item{${\bf u_1} \in {\cal P}$: For these harmonic masks, the field after the Lyot stop is uniform in amplitude with a phase gradient ${\bf x_1}$. }
\item{${\bf u_1} \in {\cal Q}$:  For these masks, the field inside the Lyot stop is zero.}
\item{${\bf u_1} \not\in ({\cal P} \cup {\cal Q})$:  For these masks, the field after the Lyot stop does not have uniform amplitude.}
\end{trivlist}
These three kinds of harmonic masks correspond to the three kinds of virtual pupils
illustrated in Figure~6 of \citet{kuch02}.

A band-limited mask is defined to be a continuous sum of harmonic masks of the first variety;
\begin{equation}
\hat M({\bf x})= \int_{\bf u_1 \in P}  \, M({\bf u_1}) \, e^{2 \pi i {\bf u_1} \cdot {\bf x}} \, d{\bf u_1}.
\end{equation}
A notch filter mask is defined to be a continuous sum of harmonic masks of the first and second varieties;
\begin{equation}
\hat M({\bf x})= \int_{\bf u_1 \in (P \cup Q)}  \, M({\bf u_1}) \, e^{2 \pi i {\bf u_1} \cdot {\bf x}} \, d{\bf u_1}.
\end{equation}
Combining this expansion and Equation~\ref{eq:fharmonic} using the linear property
of $F(u)$ described above, we find that in
a coronagraph with a notch filter mask, the field amplitude after the Lyot stop is
\begin{eqnarray}
F({\bf u}) &=& \int_{\bf u_1 \in (P \cup Q)}  \, M({\bf u_1}) L({\bf u}) A({\bf u}-{\bf u_1}) e^{-2 \pi i ({\bf u}-{\bf u_1}) \cdot {\bf x_1}} \, d{\bf u_1} \\
&=& \int_{\bf u_1 \in P}  \, M({\bf u_1}) L({\bf u}) e^{-2 \pi i ({\bf u}-{\bf u_1}) \cdot {\bf x_1}} \, d{\bf u_1} \\
&=&  L({\bf u}) e^{-2 \pi i {\bf u} \cdot {\bf x_1}} \int_{\bf u_1 \in P}  \, M({\bf u_1}) e^{2 \pi i {\bf u_1} \cdot {\bf x_1}} \, d{\bf u_1}. \\
\end{eqnarray}
To interpret this equation, let us define the band-limited part of
$\hat M({\bf x})$ as
\begin{equation}
\hat M_{BL}({\bf x})= \int_{\bf u_1 \in P}  \, M({\bf u_1}) \, e^{2 \pi i {\bf u_1} \cdot {\bf x}} \, d{\bf u_1}.
\end{equation}
Now we can write
\begin{equation}
F({\bf u}) = \hat M_{BL}({\bf x_1}) L({\bf u}) e^{-2 \pi i {\bf u} \cdot {\bf x_1}}
\qquad \mbox{for a notch filter mask.}
\end{equation}
The final image field is the Fourier transform of this quantity,
$\hat F({\bf x})=\hat M_{BL}({\bf x_1}) \hat L({\bf x - x_1})$, and
the final image intensity is the absolute square of the Fourier transform of this quantity,
\begin{equation}
|\hat F({\bf x})|^2 = |\hat M_{BL}({\bf x_1})|^2 |\hat L({\bf x - x_1})|^2
\qquad \mbox{for a notch filter mask.}
\end{equation}
In other words, for a notch filter mask, the PSF shape is $|\hat L({\bf x})|^2$,
independent of ${\bf x_1}$, the position of the source on the sky.  The PSF is attenuated by
a factor $|\hat M_{BL}({\bf x_1})|^2$, the amplitude transmissivity of the band-limited
part of the mask evaluated at the source position.  The band-limited part of a notch filter
mask can generally be found by applying a low-pass filter to the mask function.

\end{document}